\documentclass[12pt]{iopart}

\usepackage{placeins}
\usepackage{iopams}
\usepackage{graphicx}
\usepackage{hyperref}
\usepackage{epstopdf}

\newcommand{\cfrac}[3][c]{{\displaystyle\frac{%
  \strut\ifx r#1\hfill\fi#2\ifx l#1\hfill\fi}{#3}}%
  \kern-\nulldelimiterspace}

\renewcommand\Re{\mathop{\mathrm{Re}}}

\begin{document}

\title{Stability of rapidly-rotating charged black holes in $AdS_5 \times S^5$ }

\author{Micha Berkooz$^1$, Anna Frishman$^1$ and Amir Zait$^1$}

\eads{\mailto{micha.berkooz@weizmann.ac.il}}

\address{
$^1$Department of Particle Physics and Astrophysics,
The Weizmann Institute of Science,
Rehovot 76100, Israel
}

\begin{abstract}
We study the stability of charged rotating black holes in a consistent truncation of
Type $IIB$ Supergravity on $AdS_5 \times S^5$ that degenerate to extremal black holes with zero entropy.
These black holes have scaling properties between charge
and angular momentum similar to those of Fermi surface-like operators in a
subsector of ${\cal N}=4$ SYM.
By solving the equation of motion for a massless scalar field in this background,
using matched asymptotic expansion followed by a numerical solution scheme,
we are able to compute its Quasi-Normal modes, and analyze it's regime of (in)stability.
We find that the black hole is unstable when its
angular velocity with respect to the horizon exceeds 1 (in units of $1/l_{AdS}$).
A study of the relevant thermodynamic Hessian reveals a
local thermodynamic instability which occurs at the same region of parameter space. We comment on the endpoints of this instability.
\end{abstract}

\maketitle

\section{Introduction}
The AdS/CFT correspondence, first formulated by Maldacena \cite{Maldacena1998}, has been used to study the microscopic degrees of freedom of black holes, which are known to possess a large entropy.
Indeed, significant progress was made in understanding detailed properties of supersymmetric (SUSY) black holes, via their CFT duals. However, the case of little or no supersymmetry is much less understood, and it is not always possible to pinpoint the microscopic duals of black holes.

The most studied example of $AdS/CFT$ correspondence is the duality between type $IIB$ string theory on $AdS_5 \times S^5$ and ${\cal N}=4$ $SU(N)$ $SYM$, and it is also the one we have in mind in this paper. We will be interested in black holes that have angular momentum in a two-plane,
which we denote by $J_{\phi}$, and two equal charges $Q_1=Q_2 \equiv Q$. Both the angular momentum in the transverse two-plane and the third $U(1)$ charge are set to zero. The extremal limit of these black holes has zero entropy. For a specific one dimensonal set of $Q$ and $J_\phi$ the extremeal limit has 1/16 of SUSY preserved, but other than this measure zero set, no SUSY is restored.  

The fact that these black holes have zero entropy, and the details of the scaling relations between angular momentm and charge (in the large charges limit) makes them ideal candidates for Fermi surfaces in the dual ${\cal N}=4$ superconformal field theory.  Indeed, in \cite{Berkooz2012} we made a conjecture regarding a specific CFT dual configuration, relying on the structure of closed subsectors - the $PSU(1,1|2)$ sector - of the dual theory. 
In order to support the conjecture, we studied the spectrum of scalar perturbations,
and it was shown to match our expectations when the black holes are fast rotating and close to extremality. In some more detail, the appropriate dual was conjectured to be an effectively 1+1 chiral Fermi surface, and the spectrum of the quasi-normal modes of the scalar above matched the expected dimension from the CFT. 

Further understanding of the proposed duality requires a better understanding of the phase diagram on both sides. This paper presents an analysis of the stability, or more precisely instability, of these black hole backgrounds, and shows that they are prone to two types of instabilities - superradiant instability and thermodynamic instability - in part of the parameter space of $J_\phi$ and $Q$. We work in the double limit of large charge/angular momentum and vanishing temperature, taking first the charge to infinity and only then the temperature to zero. 
In addition, we discuss the near-horizon limit of these black holes. The extremal black hole has a pinching $AdS_3$ orbifold near horizon, as discussed in \cite{Boer2011, Sheikh-Jabbari2011}, where it was shown to appear in the extreme Kerr black hole in $AdS_5$, when one of the angular momenta is set to zero.

To check for the existence of a superradiant instability we add a
massless scalar to the black hole background and perform a linear stability analysis.
This scalar, which is the 10D axion-dilaton, can be added within a consistent truncation of the type $IIB$ theory on $S^5$. We compute the Quasi-Normal Modes (QNMs) of the scalar in the black hole background numerically for different values of parameters in the above-mentioned limit.

To the best of our knowledge, linear stability analysis has not been performed for these black holes.
However, there have been several studies aimed at computing the scalar
field QNMs of rotating black holes in $AdS$.
Analytical results have been obtained, but only for slow-rotating small black holes (as compared to the AdS radius), where one can neglect the effect of the black hole far away from it. The analysis was carried out in $D=4$ \cite{Cardoso2004} and in $D=5$ \cite{Aliev2008}. The fast rotating large black hole regime has proven more difficult to explore, even numerically \cite{Uchikata2009}.

At the linear level, the equation of motion for the added scalar is that of a minimally coupled massless scalar field.
The equation is separable, reducing it from a single PDE to a pair of coupled ODEs, a radial and an angular one. For a generic choice of the black hole's parameters, these are
both Heun equations, i.e., second order regular differential equations having four regular singular points.
In the fast-rotation limit, two of the singularities in the angular equation merge.
Then, using a matched asymptotic analysis developed in \cite{Lay1999,Berkooz2012},
with some modifications, we are able to solve this equation.
The solution yields the separation constant in terms of the complex frequency of the scalar.
Finally, we use the Continued Fraction Method (CFM) to solve the radial equation numerically and find eigenfrequencies.

It was shown in \cite{Hawking2000} that for Kerr-AdS black holes there can be no superradiance in $AdS_5$ if the angular velocity at the horizon, denoted $\Omega_{\phi}$, is smaller than one. For $\Omega_{\phi}>1$, on the other hand, one expects the appearance of supperradiance, which in an $AdS$ background renders the black hole unstable.
 Our expectation is that this remains true in the large charge/angular momentum limit, so that the black hole becomes superradiant at $\Omega_{\phi} > 1$. However, in the large charge/large angular momentum limit of our black holes, the angular momentum is parametrically larger than the charge and $\Omega\rightarrow 1$ for all black holes in this limit. We therefore need to examine more closely from which side it approaches $1$, as a function of the subleading charge.
 
Parameterizing by $\alpha$ the ratio between the angular momentum and the charge squared, $\alpha=\frac{N^2 J_{\phi}}{Q^2}$ we have, close enough to extremality, $\Omega_{\phi}<1$ when $\alpha<2$ and $\Omega_{\phi}>1$ when $\alpha>2$. For $\alpha=2$, $\Omega_{\phi}=1$ and the extremal black hole satisfies a SUSY BPS bound, i.e., the SUSY BPS black holes divides the parameter space into two regions with very distinct features, with expected stability on one side and instability on the other. Indeed, our numerical analysis verifies this picture and we find no superradiant instability for $\alpha \leq 2$. For $\alpha>2$ instability is possible only for modes with very large values of angular momentum, $m_{\phi} \approx \sqrt{J_{\phi}/{N^2}}$ of the scalar.
For such values, the perturbative expansion we used to obtain the solution is under lesser control. However, the application of the solution in this case displays superradiance accompanied by an instability.

Next we turn to the issue of a thermodynamic instability. To determine wether the black hole is thermodynamically stable we have analyzed the eigenvalues of the thermodynamic Hessian. For $\Omega_{\phi} < 1$ we find that the black hole is thermodynamically stable,
i.e. all eigenvalues of the Hessian are positive .
For $\Omega_{\phi}>1$ there is one negative eigenvalue corresponding to the unstable direction, characterized by an equal change in energy and angular momentum and no change in the charge.
We thus find that both thermodynamic and superradiant instabilities are present for
black holes with $\Omega_{\phi}>1$, and no instability is observed for black holes with $\Omega_{\phi} \le 1$. This is reminiscent of the Gubser-Mitra conjecture, but for rotating black holes \cite{Gubser2001}. We will return to this point in the summary and discussion section.

The outline of this paper is as follows.
In section \ref{stability_section} we briefly review the stability problem for black holes,
specifically focusing on  the phenomenon of superradiance and superradiant instability.
Section \ref{black_hole_section} is dedicated to describing the black hole background and examining its extremal limit, followed by a discussion of the near-horizon limit of such geometries and a short description of the CFT dual.
In section \ref{QNM_Sec} we solve the angular equation using an asymptotic matching technique,
and set up a numerical procedure for solving the radial equation using the Continued Fraction Method.
The results of the computation are presented in section \ref{QNM_res_section},
where we compute several Quasinormal modes using the method described here.
In section \ref{thermodynamic_section}
we analyze the thermodynamic instabilities of the rapidly rotating black hole.
Finally, the results, along with possible end-points for the instabilities, are discussed in section \ref{discussion}.

\section{Black hole stability}
\label{stability_section}
The stability of black holes is the subject of extensive research.
Usually, the question of stability is divided into two parts: classical stability
and quantum stability.
The former can be studied in general relativity, usually in linear perturbation
theory with higher orders for zero modes.
The latter can be understood using semiclassical methods such as black hole thermodynamics.
In this paper, we discuss both types of (in)stability, with emphasis on classical stability.

\subsection{Classical Stability}
One usually studies the classical stability
of black hole backgrounds by using the linearized equations of motion,
within the low-energy effective action, for whatever fields are present in the theory.
While some works have attempted to solve the full non-linear stability problem and find
the final stable configuration, this is usually a very difficult task which
we shall not discuss here.

In order to analyze the linear stability of black holes,
one first writes the linearized equations
of motion around the black hole background for the different fields present in the (super)gravity background. Typically, this system is not tractable analytically,
or numerically (with any reasonable amount of effort).
However, in some cases, the various equations decouple,
so that the stability can be assessed for each field (or combination of fields) separately.
As black holes are time independent, one can then look for the eigenmodes of time translation for each
field on this background.
Since we are dealing with a dissipative system, as all waves have to be incoming at the event horizon,
these are not normal modes, and are rather called Quasi-Normal Modes
(QNMs for short), having complex eigenvalues under the time translation symmetry (for a review of QNMs and their role in black hole physics see \cite{Berti2009}).
Stability follows directly from the sign of the imaginary part of these eigenvalues - in our convention, when the sign is positive, the perturbation decays, and when it is negative, the perturbation grows, signaling an instability.

One of the most well-studied instabilities of this type is the superradiant instability, occurring in charged or rotating black holes. The black holes discussed in this paper are both charged and rotating,
and hence this phenomenon merits special attention.

\subsubsection{Superradiant Instability}
\label{superradiance_section}
The phenomenon of superradiance allows the extraction of energy from a black hole by scattering specific modes, whose reflection coefficient from the event horizon is larger than one. It was first discovered by Zel'dovich \cite{Zel1971}, who noticed that a
conducting cylinder rotating about its axis with angular velocity $\Omega$ can amplify electromagnetic
modes impinging on it which satisfy $\omega - m \Omega < 0$, $m$ being the angular quantum number and $\omega$ the eigenmode's frequency.
This process thus extracts rotational energy from the cylinder.
There have been many results identifying superradiance in different types of black holes:
\begin{itemize}
\item Bekenstein \cite{Bekenstein1973} has shown that if the area theorem holds then modes satisfying
    $\Re\omega < m_{\phi} \Omega_{\phi}$ will lead to the extraction of energy from the black hole.
    Here, $m_\phi$ is the conserved quantum number of the azimuthal angle $\phi$,
    with respect to a non-rotating frame at infinity.
    Also, we denote by $\Omega_{\phi}$ the angular velocity of the black hole at the event horizon,
    defined as the coefficient of $\partial_{\phi}$ in the killing vector which becomes null on the horizon,
    $l = \frac{\partial}{\partial\tau} + \Omega_{\phi} \frac{\partial}{\partial{\phi}}$.
    It was then shown by direct computation that indeed scalar field modes satisfying $\Re \omega < m_{\phi} \Omega_{\phi}$ impinging on the black hole are superradiant \cite{Starob1973}.
    The same can be shown for gravitational and electromagnetic perturbations \cite{Teukolsky1974}.
\item Consequentially, it can be shown that given boundary conditions
    that correspond to reflecting the superradiant modes back onto the black hole,
    say by placing it in a box,  it will grow unstable.
    This is referred to in the literature as the 'Black Hole Bomb' \cite{Cardoso2004a}.
    One thus expects black holes in an AdS background with large enough angular velocity to be unstable due to superradiance.
\item Indeed, one can also show by explicit computation
    that in small slowly-rotating Kerr-AdS black holes in $Ads_4$
    a scalar field mode is unstable iff it is superradiant \cite{Cardoso2004}.
    This result has been obtained for gravitational perturbations as well \cite{Cardoso2006}. These solutions have been obtained by solving the angular equation perturbatively about the flat-space solutions, using the slow rotation and small size of the black holes.
    Similar analysis has been performed for charged rotating $AdS_5$ black holes \cite{Aliev2008}. When these conditions are not satisfied, an analytic solution is typically not available, yet several numerical methods have been used to find QNMs, displaying instabilities in some cases (e.g. \cite{Uchikata2009}).
\end{itemize}

In this paper we explore black holes in an asymptotically $AdS_5$ spacetime,
which are rotating with angular momentum scaled to $\infty$
(around one of the rotation planes).
The size of these black holes is comparable to $l_{ads}$.
For such black holes, there are no known analytical solutions for the QNMs.
There has been an attempt to solve the linearized equation of motion numerically
for the Kerr-Ads black hole, and even in this case one encounters problems in the fast-rotation regime
\cite{Uchikata2009}. Here, we apply a technique introduced in \cite{Lay1999} and refined in \cite{Berkooz2012} to find semi-analytically the QNMs for fast-rotating black holes in $AdS_5$.

\subsection{Semiclassical Instabilities}
Black hole thermodynamics goes back to Bekenstein's and
Hawking's seminal works \cite{Bekenstein1973b, Hawking1975}.
The thermodynamic effects in black hole backgrounds can be studied using
semiclassical calculations. There are two types of instabilities which are discussed in this context:
\begin{itemize}
\item A gravitational solution may not be a local minimum of the free energy, in which case it is thermodynamically unstable. The prime example of this is the Schwarzschild black hole.
    A generalization of this, for multi-charged black holes in various dimensions has also been studied (e.g. \cite{Davies1977}). Specifically, such thermodynamic instabilities manifest themselves as non-positive modes of the thermodynamic Hessian
    \begin{equation}
    H_{m n} = -\frac{\partial^2 S}{\partial x^m \partial x^n},
    \end{equation}
    where $x_m$ are the extensive variables describing the black holes,
    for example $E$, the total energy, $Q$, the electric charge and $J$, the angular momentum.
    While this computation is straightforward in simple backgrounds, such as Schwarzschild and Kerr, it becomes more complex as charges and angular momenta are added in more complicated gravity theories.
\item Global instabilities: these occur when at least two gravitational backgrounds with the same charges exist, and are local minima of the free energy. In this case, the relevant free energy may be lower for one of them compared to the other, rendering it the stable configuration, with the other being metastable. If, in some region of the parameter space, these roles are reversed, a phase transition will occur. The most well known example of this phenomenon is the Hawking-Page phase transition \cite{Hawking1983} which has been given a beautiful interpenetration within the framework of AdS/CFT, relating it to the confining/deconfining phase transition in the dual gauge theory \cite{Witten:1998zw}.
\end{itemize}

\section{The Black Hole}
\label{black_hole_section}
\subsection{Description of the black hole solution}
The notations in this paper follow closely those of \cite{Mei2007}.
We work within the consistent truncation of type
$IIB$ Supergravity on $AdS_5 \times S^5$ described in \cite{cvetic}.
The field content comprises of the metric, two neutral scalars and three abelian $U(1)$ vector fields.
It can also be consistently coupled to a dilaton and an axion, which are massless, as was discussed in  \cite{cvetic} and later used in \cite{Berkooz2012}.
The bosonic part of the supergravity Lagrangian is given by
\begin{eqnarray}
\label{lagrangian}
{\cal L} = & \sqrt{-g} \left[ R - \frac{1}{2} \sum_{\alpha=1}^{2}{(\partial \varphi_{\alpha})^2} + \sum_{i=1}^{3}{\left(4g^2 X_i^{-1} - \frac{1}{4} X_i^{-2} {\cal F}^i_{\mu \nu} \mathcal{F}^{i \mu \nu}\right)}\right] + \nonumber \\
& \frac{1}{24} |\epsilon_{ijk}| \epsilon^{uv \rho \sigma} \mathcal{F}^i_{uv} \mathcal{F}^j_{\rho \sigma} A_{\lambda}^k
\end{eqnarray}
Here,  $l_{ads} \equiv l =\frac{1}{g}$ is the AdS radius, $A^i$ are the three $U(1)$ gauge fields, and
$X_i$ are the three uncharged scalars, constrained by $X_1 X_2 X_3 = 1$.
These scalars may be parameterized by
\begin{eqnarray}
X_1 = e^{-\frac{1}{\sqrt{6}}\phi_1 -\frac{1}{\sqrt{2}} \phi_2},\ \ \ \
X_2 = e^{-\frac{1}{\sqrt{6}}\phi_1 +\frac{1}{\sqrt{2}} \phi_2},\ \ \ \
X_3 = e^{\frac{2}{\sqrt{6}}\phi_1}
\end{eqnarray}
We discuss the black hole solutions that first appeared in \cite{pope1},
while following the notations of \cite{Mei2007} which contains a generalization of these solutions.
The solutions are parameterized by $\delta_1$, $\delta_3$, $m$, $a$, $b$, which map to $Q_1=Q_2$, $Q_3$, $E$, $J_L$ and $J_R$.
Motivated by dual field-theoretic considerations first presented in
\cite{Berkooz2012}, which we describe briefly in section \ref{field_theory_dual_sec},
we are interested in solutions for which $Q_3 = 0$, $J_{\psi} = 0$ and $Q_1=Q_2$.
This amounts to setting $b = 0$ and $\delta_3 = 0$.
The solution then reduces to the following metric
\begin{eqnarray}	
\label{metric2}
ds^2 = H_1^{2/3} \Biggl\{&  \left(x^2 + y^2\right) \left(\frac{dx^2}{X} + \frac{dy^2}{Y} \right) -
			\frac{ X \left( dt - y^2 d \sigma \right)^2}{\left(x^2 + y^2\right) H_1^2} + \Biggr. \nonumber \\
&\Biggl. \frac{Y \left[dt + \left(x^2 + 2 m s_1^2\right) d\sigma \right]^2}{\left(x^2 + y^2\right) H_1^2}  + y^2 x^2 d\chi^2 \Biggr\},
\end{eqnarray}
were we define $s_1 = sinh\left(\delta_1\right)$ (and $c_1 = cosh\left(\delta_1\right)$ for later use).
The functions used in the metric are
\begin{eqnarray}
X &= -2m + \left(a^2 + x^2\right) + g^2\left(a^2 + 2m s_1^2 +x^2\right) \left( 2m s_1^2 + x^2 \right) \\
Y &= \left(a^2 - y^2\right) \left( 1 - g^2 y^2\right) \\
H_1 &= 1 + \frac{2 m s_1^2}{x^2+y^2}.
\end{eqnarray}

The linear transformation to the coordinate frame that is non rotating at infinity is given by
\begin{eqnarray}
t &= \frac{1}{\Sigma_a} \tau - \frac{a}{\Sigma_a} \phi, \\
\sigma &= \frac{g^2}{\Sigma_a} \tau - \frac{1}{a \Sigma_a} \phi, \\
\chi &= \frac{1}{a} \psi,
\label{angular_trans}
\end{eqnarray}
with $\Sigma_a = 1-g^2 a^2$.
Finally, it is occasionally convenient to use an additional transformation in the non-rotating frame
\begin{eqnarray}
& x^2 =r^2-\frac{4}{3} m s_1^2 \\
& y^2=a^2 cos^2\theta.
\end{eqnarray}
The event horizon is now located at $r=r_0$,
the largest root of
\begin{equation}
\label{delta_r}
\Delta(r)=\frac{x^2 X}{r^2}.
\end{equation}

The angular velocity around the $\phi$ direction at the event horizon,
with respect to the asymptotically non-rotating AdS space, is given by
\begin{equation}
\Omega_{\phi} = \frac{a\left(1+g^2\left(r_0^2 + \frac{2 m s_1^2}{3}\right)\right)}{a^2+r_0^2 + \frac{2 m s_1^2}{3}}
\end{equation}
Conversion to natural field-theoretic units, where $Q_1=Q_2 \equiv Q$ and $Q_3$ are the cartans of $SO(6)$ and $J_{\phi},J_{\psi}$ are half-integer quantized, is performed using the relation \cite{Aharony2000}
\begin{equation}
\frac{\pi l_{ads}^3}{4G_5} = \frac{N^2}{2}.
\end{equation}
In terms of these units, the black hole's charges are \cite{Mei2007}
\begin{eqnarray}
\label{charges}
J_{\phi} &= \frac{\pi}{4 G_5} \frac{2 m a \left(1+ s_1^2\right)} {{E_{a}}^2} =N^2 g^3 \frac{ m a \left(1+ s_1^2\right)} {{E_{a}}^2}  \\
Q_{1} &= Q_{2} \equiv Q = \frac{\pi}{4 G_5} \frac{2 m s_{1} c_{1}}{E_{a}}=N^2 g^3 \frac{ m s_{1} c_{1}}{ E_{a}}  \\
E &= \frac{\pi}{4 G_5} \frac{m [ (2(g^4 a^4 + \Sigma_a + 1) + g^2 a^2 (\Sigma_a -2))s_1^2 + \Sigma_a  + 2] }{\Sigma_a^2} \\
S &= \frac{\pi^2}{4 G_5} \frac{2 \sqrt{r_0^2 - \frac{4}{3} m s_1^2} (a^2 + \frac{2}{3} m s_1^2 + r_0^2)}{\Sigma_a} \nonumber \\ 
    & = N^2 g^3 \pi \frac{ \sqrt{r_0^2 - \frac{4}{3} m s_1^2} (a^2 + \frac{2}{3} m s_1^2 + r_0^2)}{\Sigma_a}  \\
T &= \frac{r_0\left(\partial _r\Delta \right)\sqrt{r_0^2-\frac{4}{3}m s_1{}^2}} 
     {4\pi \left(a^2+r_0^2+\frac{2}{3}m s_1{}^2\right)\left(r_0^2-\frac{4}{3}m s_1{}^2\right)}. 
\end{eqnarray}
Here, $r_0$ is again the radius at which the event horizon is located.
In the rest of this paper, we set $l=1/g=1$. One can restore these factors using dimensional analysis.

\subsection{The Near Extreme and the Fast Rotation Limits}
\label{extremal_limit_sec}
First, we briefly analyze the extremal limit of the black hole.
In order to find the location of the horizon we solve $\Delta(r) = 0$ in (\ref{delta_r}).
There are two solutions for $\Delta = 0$, given by $X = 0$ or $x^2 = 0$:
\begin{eqnarray}
x^2=0:\  & r_1^2 = \frac{4m s_1^2}{3} \\
X =0:\  &r_2^2 = \frac{-1 - a^2 - \frac{4 m s_1^2}{3} + \sqrt{(a^2-1)^2 + 8m (1+s_1^2)}}{2} \\
       &r_3^2 = \frac{-1 - a^2 - \frac{4 m s_1^2}{3} - \sqrt{(a^2-1)^2 + 8m (1+s_1^2)}}{2} < 0
\end{eqnarray}
There are two horizons, located at $r_1$ and $r_2$ and the extremal limit is obtained
when $r_1=r_2$ in which case we also find that the entropy is zero.
Using these definitions we write
\begin{equation}
\label{large_X}
X=(r^2-r_2^2)(r^2-r_3^2)=(x^2 + (r_1^2-r_2^2))(x^2 + (r_1^2-r_3^2)) \equiv (z-z_2)(z-z_1),
\end{equation}
where in the last line we denoted $z=x^2$, so that $z_2 \equiv r_2^2-r_1^2$
and $z_1 \equiv r_3^2-r_1^2$.
When $r_2 < r_1$ we get a naked singularity. Thus, for the non-singular black holes we can use $z_2$
as a measure of the distance of the black hole from an extremal one.
It will be convenient to use a parametrization in terms of $\alpha= \frac{J_{\phi}/N^2}{(Q/N^2)^2}$,
$a$ and $z_2$. In terms of these parameters, we have
\begin{eqnarray}
m &=  \frac{\left(a^2+z_2\right) \left(1+z_2\right)}{2}+\frac{2 a^2}{\alpha ^2}+\frac{a^3+2 a z_2}{\alpha } \\
m s_1^2 &=  \frac{ a }{\alpha} \\
z_1 &=  -\frac{4 a +\left(a^2+1+z_2\right) \alpha }{\alpha }.
\end{eqnarray}

In this parametrization, the near extremal limit for this black hole is thus $z_2 \rightarrow 0$.
We have studied this limit in \cite{Berkooz2012},
where we have found some analytic solutions for the QNMs, which are similar to those of a $1+1$ CFT.
In the class of large rotating black holes that we consider here,
there is an additional limit which facilitates solving the equation of motion, namely the fast rotation limit.
For black holes with angular momentum scaled to infinity, the distance of the rotation
parameter $a$ from its maximal value $1$ can be used to obtain some analytical results.
In this paper we focus on the large rotation limit, and thus $z_2 \gg (1-a)$.

The BPS limit of the black hole is
\begin{equation}
El=J_{\phi}+J_{\psi}+Q_1+Q_2+Q_3=J_{\phi}+2Q_1
\end{equation}
and it is satisfied at $z_2=0$ and $\alpha=2$.
We expect superradiance, and since the spacetime is asymptotically AdS also an instability, if
$\Omega_{\phi}>1$. Rewriting $\Omega_{\phi}$ in terms of $\alpha$ and $z_2$ we get
\begin{equation}
\Omega_{\phi}=\frac{a \left(2 a+\alpha +\alpha  z_2\right)}{a (2+a \alpha )+\alpha  z_2}=1+\frac{\left(2-\alpha +\alpha  z_2\right) (a-1)}{2+\alpha +\alpha  z_2}+O\left((a-1)^2\right).
\end{equation}
This expression shows that to leading order in $(1-a)$
we have $\Omega_{\phi}=1$ independently of both $\alpha$ and $z_2$.
However, the next-to-leading order correction shows that for fixed $\alpha$ there is a threshold value for the off-extremality parameter $z^{\ast}_2 = 1-2/\alpha$,
such that $\Omega_{\phi}<1$ for $z_2 > z_2^{\ast}$ and  $\Omega_{\phi}>1$ for $z_2 < z_2^{\ast}$.
Since $z_2$ is always positive, we can see that for $\alpha < 2$
we have $\Omega_{\phi} < 1$ for all values of $z_2$.
As the appearance of superradiant instability is expected only for $\Omega_{\phi} > 1$, in the limit $a \rightarrow 1$ it should appear only for $\alpha > 2$ and sufficiently low $z_2$.
Note that in the extremal limit, i.e when we set $z_2$ to zero first, the crossing point from $\Omega_{\phi}>1$ to $\Omega_{\phi}<1$ is precisely at $\alpha=2$ which corresponds to the BPS black hole.

It is useful to present the temperature and entropy in terms of $z_2$ and $\alpha$:
\begin{eqnarray}
\label{entropy_temp}
S &= N^2 \frac{2\pi \sqrt{z_2} \left(a^2+\frac{2 a}{\alpha }+z_2\right)}{\Sigma_a} \\
T &= \frac{\sqrt{z_2} \left(4 a +\alpha + a^2 \alpha +2 \alpha z_2\right)}{4 a \pi +2 \pi  \alpha  \left(a^2+z_2\right)}.
\end{eqnarray}

Let us express the appearance of superradiance in terms of the temperature. To leading order in $1-a$ we have
\begin{equation}
T=\frac{\sqrt{z_2}}{\pi}+O[1-a].
\end{equation}
There is then a critical temperature $T_c=\frac{1}{\pi}+O(1-a)$ above which $\Omega_{\phi}$ is smaller than one for any $\alpha$ and we expect no superradiant instability to develop, independently of the choice of $\alpha$.

\subsection{The Near-Horizon Limit}
It is usually useful to study the near-horizon
limit of black holes in order to understand their properties.
Such studies, mainly of extremal or near-extremal black holes, have resulted in better understanding of the black hole's microscopic degrees of freedom.
The issue of instability can also be studied using the near horizon geometry,
which is often simpler than the full geometry \cite{Durkee2010}.
Thus, it would be beneficial to find the near horizon geometry of the black holes considered here.
There is, however, an obstacle to defining this limit globally for the near extremal black hole, as we discuss below.

Examining (\ref{entropy_temp}) we see that the entropy, and thus the horizon area of the black hole, approaches zero together with the off-extremality parameter $z_2$. The extremal metric is therefore singular. In addition, computing the Ricci scalar, one can see that at $x=0$ there is a ring of singularities at $y=0$, where the $\psi$ circle pinches to a point.
This ring is hidden behind the horizon for the non-extremal black holes,
and touches the horizon when $z_2=0$.
Below we present the extremal near-horizon limit for this family of black holes.
\footnote{The analysis of this section was carried together with Joan Simon, and will be discussed further in \cite{work_in_progress_joan}}

\subsubsection*{The Extremal Near-Horizon Geometry}

{\bf $x\rightarrow 0$, fixed y}:
To take this limit carefully we first define $x=\epsilon \rho$, taking $\epsilon\rightarrow0$ in the end.
Applying this to the functions appearing in the metric we get:
\begin{eqnarray}
X &\rightarrow \left(1+ a^2+4\frac{a}{\alpha}\right) \epsilon^2 \rho^2 \\
H_1 & \rightarrow 1+\frac{2a}{\alpha y^2}.
\end{eqnarray}
If we naively take $\epsilon\rightarrow0$ we will set the time coordinate of our solution to zero.
Instead, we take $t = \frac{\hat{t}}{\epsilon}$.
We will also define $\hat{\phi} = t + \frac{2a}{\alpha} \sigma$ and $\hat{\chi} = \epsilon \chi$.
The metric then takes the form:
\begin{equation}
\fl ds^2 = H_1^{\frac{2}{3}} \Biggl\{ y^2 \Biggl(\frac{d\rho^2}{\left(1+ a^2+4\frac{a}{\alpha}\right) \rho^2} + \rho^2 d\hat{\chi}^2 - \left(1+a^2+4\frac{a}{\alpha}\right) \frac{\alpha^2}{4a^2} \rho^2 d \hat{t}^2 \Biggr) + \frac{y^2 dy^2}{Y} + \frac{Y d\hat{\phi}^2}{y^2 H_1^2} \Biggr\}.
\end{equation}
This metric has a pinching $AdS_3$ orbifold factor, as defined in \cite{Boer2010}, 
which is an $AdS_3$ except that the periodicity of the spatial circle $\hat{\chi}$ scales to zero in the this limit. 
Such geometries have been studied in the context of Extremal Vanishing Horizon (EVH) black holes in $AdS_5$ \cite{Sheikh-Jabbari2011, Boer2011}, and recently in \cite{Johnstone2013}.
It would be interesting to see whether one can use this generic structure,
along with the proposed dual, to learn more of the extremal black hole's properties and test the EVH/CFT duality in detail. 

{\bf Subsequent $y\rightarrow 0$:}
Taking now the $y \rightarrow 0$ limit yields:
\begin{equation}
\fl ds^2 = \left(\frac{2a}{\alpha}\right)^\frac{2}{3} y^{2/3}
\Biggl\{ \frac{d\rho^2}{\left(1+ a^2+4\frac{a}{\alpha}\right) \rho^2} +
\frac{dy^2}{a^2} + \rho^2 d\chi^2 - \left(1+ a^2+4\frac{a}{\alpha}\right) \frac{\alpha^2}{4a^2} \rho^2 d \hat{t}^2 + \frac{\alpha^2}{4} d\hat{\phi}^2 \Biggr\}
\end{equation}
We see that the radius of curvature of the $AdS_3$, and actually of the entire space, approaches zero when $y \rightarrow 0$. Hence $y=0$ is a naked singularity in the extremal case.

The loci $y=0$ is also singular in the following sense: the distance to $x=0$ at $y\ne 0$ is infinite in the extremal limit, whereas the distance to $x=0$ at $y=0$ is finite. The situation is exasperated by the fact that the distance to $y=0$ along the $x=0$ horizon is finite even at extremality.

The computation is straightforward.
The metric component in the radial direction is given by
\begin{equation}
g_{xx}=\left(1+\frac{2a}{\alpha\left(x^2+y^2\right)}\right)^{2/3}\frac{\left(x^2+y^2\right)}
{x^2(x^2+1+4\frac{a}{\alpha}+a^2)}
\end{equation}
Now, taking the limit $x\rightarrow0$ with finite $y$, we obtain
\begin{equation}
g_{xx}=\left(1+\frac{2a}{\alpha y^2}\right)^{2/3} \frac{y^2}
{x^2(1+4\frac{a}{\alpha}+a^2)}\sim\frac{1}{x^2}
\end{equation}
We see that the distance to the horizon, located at $x=0$, along the radial direction is infinite.
This is the familiar 'throat' of extremal rotating black holes.

On the other hand, approaching the $x=0$ limit at fixed $y=0$, the metric component in the radial direction is now given by
\begin{equation}
g_{xx}=\left(1+\frac{2a}{\alpha x^2}\right)^{2/3} \frac{1}{x^2+1+4ms_1^2+a^2}\sim\frac{1}{x^{4/3}}.
\end{equation}
Performing the integration from finite $x$ to $x=0$ now yields a finite proper distance.

\subsection{Field Theory Dual}
\label{field_theory_dual_sec}
For completeness, we would like to present the motivation for studying this class of black holes, which is
rooted in the AdS/CFT duality \cite{Maldacena1998}. In \cite{Berkooz2012}
a conjecture was made regarding the field theory dual of these black holes.
The dual conformal field theory for this supergravity theory is ${\cal N}=4$
$SU(N)$ Super Yang-Mills (SYM).
In this theory, there are several known sectors which are closed under the dilatation operator, which is the radial quantization equivalent of the hamiltonian \cite{Beisert2004}. Thus, a state composed of the operators in such a sector will only mix with other states within the sector. One of these sectors, named the $PSU(1,1|2)$ sector,
contains only partons with the following relations between the charges
\begin{equation}
\label{psu_constraint}
\Delta_0 = 2 J_L + \hat{Q}_1 + \hat{Q}_2 + \hat{Q}_3 = 2J_R + \hat{Q}_1 + \hat{Q}_2 - \hat{Q}_3
\end{equation}
where $\Delta_0$ is the classical scaling dimension, $J_L$ and $J_R$ are the $SU(2) \times SU(2)$ quantum numbers, and
$\hat{Q}_1$, $\hat{Q}_2$ and $\hat{Q}_3$ are the $SU(4)$ $R$-charges spanning the $U(1)^3$ Cartan subalgebra of $SO(6)\cong SU(4)$, which are, of course, the $U(1)^3$ charges discussed before in the $AdS_5 \times S^5$ supergravity dual.

The ${\cal N}=4$ $SU(N)$ SYM theory contains a single supermultiplet in the adjoint of the $SU(N)$ gauge symmetry. This multiplet includes a gauge field $A_{\mu}$, a fermion $\psi_a$ and its conjugate $\bar{\psi}^a$, in the fundamental and anti-fundamental representations of the $SU(4)$ R-symmetry, respectively, and a scalar $\phi_{[ ab ]}$ in the ${\bf 6}$ of $SU(4)$. The lower indices on these fields are fundamental indices of the $SU(4)$ $R$ symmetry, while upper indices are anti-fundamental ones.

The constraints (\ref{psu_constraint}) retain only four types of partons from the full set
\begin{eqnarray}
\label{CFT_fields}
\phi_1^{(n)} &\equiv \mathcal{D}^n_{1\dot{1}} \phi_{24} & \phi_2^{(n)} \equiv \mathcal{D}^n_{1\dot{1}} \phi_{34} \nonumber \\
\psi^{(n)} &\equiv \mathcal{D}^n_{1\dot{1}} \psi_{14} & \bar{\psi}^{(n)} \equiv \mathcal{D}^n_{1\dot{1}} \bar{\psi}^1_1.
\end{eqnarray}
Here, $D_{1\dot{1}}$ is the covariant derivative $D_{\alpha \dot{\alpha}}$ (exchanging the $SO(4)$ of Euclidean rotation with $SU(2) \times SU(2)$ notation) with $\alpha = \dot{\alpha} = 1$, and $n$ indicates the number of derivatives operating on the fields.

Each of these fields also sits in the adjoint of the gauged $SU(N)$ so it has an additional adjoint index which we have so far suppressed. We will sometimes wish to write this index explicitly, e.g. $\phi_{(j)}^{n}$ where $j=1...dimG=N^2-1$.
One can build gauge invariant quantities in the following way.
Given an operator $\Psi$ in the adjoint of $SU(N)$, define $Jdet[\Psi]$ as
\begin{eqnarray}
Jdet[\Psi] = \epsilon^{a_1 a_2 ... a_g} \Psi_{(a_1)} \Psi_{(a_2)} ... \Psi_{(a_g)} && \Psi = \sum_{a=1}^{g} \Psi_{(a)} T^a\ (g=dimG).
\end{eqnarray}
The charges of the fermions in this sector are given in the following table.
\begin{table}[ht!]
\caption{Quantum numbers for the fermionic content of the \textit{PSU(1,1$|$2)} subsector, $j_{p}$ refers to the angular momentum of the $SU(2)_p$ automorphism.}
\begin{indented}
\item[] \begin{tabular}{@{}ccccccccc}
   \br
   & $j_L$ & $j_R$ & $Q_1$ & $Q_2$ & $Q_3$ & $j_{p}$ \\
   \mr
  $\psi^{(n)}$ & $\left(n+1\right)/2$ & n/2 & 1/2 & 1/2 & 1/2 & 1/2  \\
  $\bar{\psi}^{(n)}$ & n/2 & $\left(n+1\right)/2$ & 1/2 & 1/2 & -1/2 & -1/2  \\
  \br
\end{tabular}
\end{indented}
\end{table}

The pair of fermions in \ref{CFT_fields} enjoy a further $SU(2)$ automorphism, under which $\psi^{(n)}$ and $\bar{\psi}^{(n)}$ transform as a doublet, and we will denote them $\Psi_1^{(n)}$ and $\Psi_2^{(n)}$, respectively, to make the $SU(2)$ action manifest. This symmetry can be employed in order to build states which cannot mix with any other state under the dilatation operator. One can build such states which have the same quantum numbers as the black holes we examine. This has been done in detail in \cite{Berkooz2012}, where states resembling a Fermi surface have been constructed. These satisfy the field theory BPS bound at zero coupling, and are conjectured to be duals of the extremal BPS black holes within this family, i.e. those having $\alpha=2$.

Using the notations described before, the basic operator, corresponding to the $\alpha=2$ extremal black hole, is defined as
\begin{equation}
\label{operator}
\mathcal{O}^{(K)} = Sym\left[\prod_{n=0}^{K-1} Jdet\left[\Psi_2^{(n)}\right] \prod_{m=K}^{2K-1} Jdet\left[\Psi_1^{(m)}\right]\right].
\end{equation}
Here, $Sym[]$ stands for a symmetrization of the operator with respect to the $SU(2)$ doublet indices $\{1,2\}$, placing the operator in the highest $SU(2)$ pseudo-spin state,
with $J^2 = 2K N^2(2K N^2+1)$ and $J_z=0$.
Higher values of alpha can be obtained by removing some of the operators in $\mathcal{O}^{(K)}$, thus puncturing holes in the Fermi surface. While there may be several ways to obtain the same set of charges, the entropy is still $O(1)$ and not $O(N^2)$, so it cannot be seen on the gravity side.

\section{Scalar Perturbations in the Black Hole Background}
\label{QNM_Sec}
In this paper we are interested in near extremal, rapidly rotating black holes such that $\Sigma_a \equiv (1-a^2) \ll \sqrt{z_2}$ which results in a black hole with temperature close to zero, but non-zero entropy.
We wish to examine the possibility of superradiant instability
in these black holes and extract information about the QNM of perturbations on this background.

In order to be able to interpret the resulting perturbation spectrum in the dual CFT,
we would like for the perturbing field to be part of the consistent truncation
 of the full type $IIB$ theory.
While there are two scalars in the Lagrangian (\ref{lagrangian}), they are both massive, and their equation of motion is not separable. However, as shown in \cite{Cvetic2000}, it is possible to use the symmetries of the ten-dimensional theory to extend the consistent truncation considered thus far,
allowing us to include the axion and dilaton, which transform under the $SL(2,\mathbb{R})$ symmetry of $IIB$.

At the linearized level, the equation of motion for the dilaton
decouples and turns out to be that of a minimally coupled massless uncharged scalar.
Thus, this is the simplest field we can consistently turn on to check the existence of an instability.
The equation of motion is given by
\begin{equation}
\frac{1}{\sqrt{-g}}\partial_{\mu} \sqrt{-g} g^{\mu \nu} \partial_{\nu} \Phi = 0
\end{equation}
Plugging in the ansatz
\begin{equation}
\Phi(x,y,\tau,\phi,\psi) = e^{-i w \tau + i m_{\phi} \phi +i m_{\psi} \psi} S(y)R(x),
\end{equation}
the equation of motion can be separated into two ODEs.
This separation of variables introduces a new parameter, $\lambda$, in addition to $m_{\phi}$, $m_{\psi}$ and $\omega$.
We will carry out the analysis of stability in the limit limit $\Sigma_a \rightarrow 0$
with $\Sigma_a \ll \sqrt{z_2}$.
Some aspects of the analysis are similar to that of \cite{Berkooz2012}.
However, there the opposite limit was considered,
which in some sense 'reverses' the roles of the angular and radial equations.

The outline of this section is the following.
In section \ref{angular_eq_sec} we examine the angular equation, i.e. the ODE with respect to $y$.
We then proceed to solving it using a matched asymptotic procedure first described in \cite{Lay1999}, with some modifications, in section \ref{asymptotic_expansion_sec}.
In addition, we extend the analysis to next-to-leading order.
The result of the analysis is a relation between the separation parameter $\lambda$ and the parameters $m_{\phi}$, $m_{\psi}$ and $\omega$,
along with a new integer $n$.
In section \ref{radial_eq_sec} we apply the Continued Fraction Method to the radial equation, resulting in a discrete
spectrum for the Quasinormal modes, which we compute numerically.

\subsection{The Angular Equation}
\label{angular_eq_sec}
The equation, depending only on the angular coordinate $y$, is
\begin{equation}
\fl \frac{1}{ y }\partial _y\left(y Y \partial_yS(y)\right) - 
\left( \frac{\left(y^2 \left(a m_{\phi }- \omega \right)-am_{\phi }+a^2\omega \right)^2}{Y}+\frac{a^2 m_{\psi}^2}{ y^2 }+\lambda \right) S(y)=0.
\end{equation}
If one uses the variable $u=\frac{a^2-y^2}{a^2}$, the equation has four regular singular points.
 Thus, it can be brought to the form of a Heun equation
\begin{equation}
\label{heun eq}
F''(u)+\left(
\frac{\gamma}{u}+ \frac{\delta}{u-1}  +\frac{\epsilon}{u-s}\right)F'(u)+ \frac{(\alpha_H  \beta_H u-q)}{u(u-s)(u-1)}F(u)=0,
\end{equation}
with the four regular singular points located at $r = 0,1,s,\infty$,
and $\epsilon = (\alpha_H +\beta_H -\gamma -\delta +1)$.
The transformation to the canonical form of Heun's equation is performed using the following ansatz
\begin{equation}
S(u)=(u-1)^{\mu }u^{\eta }\left(u+\frac{1}{a^2}-1\right)^{\chi }F(u)
\end{equation}
where the newly added constants are chosen so that
they cancel the higher order poles in the coefficient of $F(u)$.
Without loss of generality, we can use the following values for them:
\begin{equation}
\eta =  \frac{|m_{\phi}|}{2} \qquad \mu  =  \frac{|m_{\psi}|}{2} \qquad \chi  = -\frac{\omega }{2}.
\end{equation}
The additional regular singular point, $s$, is given in terms of $a$ by
\begin{equation}
s = 1-\frac{1}{a^2} < 0.
\end{equation}
The other parameters of the Heun equation are given by
\begin{eqnarray}
\alpha_H = & 2+ \mu +\eta +\chi  =  \frac{1}{2} \left(4 + |m_{\phi }|+|m_{\psi }|- \omega \right)  \\
\beta_H = & \mu +\eta +\chi  =  \frac{1}{2} \left(|m_{\phi }|+|m_{\psi }|- \omega \right)  \\
\gamma = & 1+2\eta  =  1+|m_{\phi }|  \\
\delta = & 1+2\mu  =  1+|m_{\psi }|   \\
\epsilon = & 1+2\chi =1- \omega   \\
q = & \frac{ \omega  m_{\phi }}{2 a}+\frac{1}{4} \left(-2  \omega +4 \left|m_{\phi }\right|+2 - \omega  |m_{\phi }|+2 |m_{\psi }|+2 |m_{\phi }| |m_{\psi }|+m_{\phi }^2\right)+ 
\nonumber  \\
& -\frac{\left(\lambda +2 \left|m_{\phi }\right|+2 |m_{\psi }|+2 |m_{\phi } m_{\psi }|+m_{\phi }^2+m_{\psi }^2\right)}{4 a^2}.
\end{eqnarray}
The physical region $y \in [0, a]$ corresponds to
the canonical interval $u \in [0,1]$, with $y=0$ corresponding to $u = 1$ and $y=a$ to $u=0$.
The point $y=1$ corresponds to $u=s<0$.
In the limit $s \rightarrow 0$ two singular regular points merge, and the equation
becomes hypergeometric.
As a result, we can use an asymptotic matching technique to solve the equation.

\subsection{Matched Asymptotic Expansion for the Angular Equation}
\label{asymptotic_expansion_sec}
We will follow the method of \cite{Lay1999}, refined in \cite{Berkooz2012}.
However, some modifications are required, since the Heun parameters here are no longer constant, but rather depend here on an unknown quantity $\omega$. They also contain the parameter $|s|$ itself, and thus must also be included in the perturbative expansion.
The asymptotic matching is performed by dividing the interval $u \in [0,1]$ into two overlapping regions,
the far region, where $u \gg |s|$, and the near region, with $u \ll 1$. The equation in each of these regions reduces to a Hypergeometric equation, and is thus solvable. The overlap region $u \sim \sqrt{|s|}$ is then used to match the two solutions, obtaining a quantization condition.
A similar method was also used in \cite{Hod2008,Hod2008a, Hod2009,Hod2010,Hod2011} to obtain analytic expressions for quasinormal modes in flat backgrounds,
where the relevant equation is a Confluent Heun Equation.

\subsubsection*{The far region $u\gg |s|$}
Here we can set $s=0$ in (\ref{heun eq})
to obtain the zeroth order equation, this amounts to setting $a = 1$
which, in terms of the Heun parameters, will affect only $q$:
\begin{equation}
q_0 \equiv q(a = 1) = \eta  (1+\eta )-\mu ^2 +\chi  (1+2 \eta +\chi )-\frac{1}{4}\lambda  -\frac{1}{4}\left(\omega -m_{\phi }\right){}^2.
\end{equation}	
After some algebra, the solution in this region is given by
\begin{equation}
F_{far}(u)=u^{\rho_1} F_{1}[\alpha_H+\rho_1,\beta_H+\rho_1,\delta,1-u].
\end{equation}
Here $F_{1}$ is the Hypergeometric function and
\begin{eqnarray}
\rho_1 &=  \frac{1}{2} \left(-\alpha_H -\beta_H +\delta +\sqrt{-4 q_0 +(-\alpha_H -\beta_H +\delta )^2}\right) \nonumber \\
& =\frac{1}{2} \left(-1 -|m_{\phi}| -2\chi +\sqrt{1+m_{\psi}^2+\lambda+(\omega-m_{\phi})^2}\right)  \\
\rho_2 &= \frac{1}{2} \left(-\alpha_H -\beta_H +\delta -\sqrt{-4 q_0 +(-\alpha_H -\beta_H +\delta )^2}\right) \nonumber \\
&= \frac{1}{2} \left(-1 -|m_{\phi}| -2\chi -\sqrt{1+m_{\psi}^2+\lambda+(\omega-m_{\phi})^2}\right)
\end{eqnarray}
satisfy the relation
\begin{equation}
\rho_1+\rho_2=\delta-\alpha_H-\beta_H.
\end{equation}
We have also used the freedom to normalize the constant
coefficient of this solution to $1$ relative to the near region solution.
\subsubsection*{The near region $u \ll 1$ }
In the near region, in order to take the limit $s\rightarrow0$ correctly, we need to first make the transformation $u= -s \xi$ in (\ref{heun eq}). Then, taking the limit $s\rightarrow0$, one obtains the zeroth order equation in this region.
The solution to this equation is given by
\begin{equation}
F_{near}(\xi)=C F\left(-\rho_1,-\rho_2,\gamma ;-\xi \right),
\end{equation}
where $C$ is a normalization constant to be fixed by the matching procedure.

\subsubsection*{Matching the two solutions }
The next step is to match the two solutions at
the point $u=\sqrt{-s}t$ with $t \sim O(1)$, expanding around $s=0$.
This is done using hypergeometric functional identities.
We have already performed the general procedure in \cite{Berkooz2012},
and obtained the following quantization condition
\small
\begin{equation}
\frac{\Gamma^2 \left[\rho _1-\rho _2\right] \Gamma \left[\gamma +\rho _2\right]\Gamma \left[-\rho _1\right]\Gamma \left[\beta_H+ \rho_2 \right]\Gamma \left[\alpha_H + \rho_2\right]}{\Gamma \left[\alpha_H +\rho _1\right]\Gamma \left[\beta_H +\rho _1\right] \Gamma^2 \left[\rho _2-\rho _1\right]\Gamma \left[\gamma +\rho _1\right]\Gamma \left[-\rho _2\right]}=\left(\sqrt{-s}\right)^{2\rho _1-2\rho _2}.
\label{zeroth}
\end{equation}
In order to extract the correct quantization condition for $\lambda$ from this equation we must check several cases of which only one leads to a true solution. Note that the elimination of other possibilities is not completely generic as there is a point in our derivation that depends on the values of the parameters of our specific equation.
We describe how to eliminate the other possibilities in Appendix A, here we focus only on the relevant case.\\
Assuming $\Re(\rho_1-\rho_2)> 0$, the RHS of (\ref{zeroth}) goes to zero as $s\rightarrow 0$ so the denominator of the LHS must diverge in this limit. This means that one of the arguments of the gamma functions in the denominator should be equal to a negative integer.
These are given, in our equation, by
\begin{eqnarray}
\label{quant_pos}
\beta_H+\rho_1 & =\mu -\frac{1}{2}+\frac{1}{2}\sqrt{1+4\mu^2 +\lambda +(\omega-m_{\phi})^2} \nonumber \\
& =  \frac{1}{2}(|m_{\psi}| -1+\sqrt{1+m_{\psi}^2 +\lambda +(\omega-m_{\phi})^2}  \\
\alpha_H+\rho_1 &= \mu +\frac{3}{2}+\frac{1}{2}\sqrt{1+4\mu^2 +\lambda +(\omega-m_{\phi})^2} \nonumber  \\
& =  \frac{1}{2}(|m_{\psi}| +3+\sqrt{1+m_{\psi}^2 +\lambda +(\omega-m_{\phi})^2}  \\
\gamma+\rho_1 &=  \frac{1}{2}+\eta-\chi+\frac{1}{2}\sqrt{1+4\mu^2 +\lambda +(\omega-m_{\phi})^2} \nonumber \\
& =  \frac{1}{2}\left(1+|m_{\phi}|-\omega + \sqrt{1+m_{\psi}^2 +\lambda +(\omega-m_{\phi})^2}\right) \\
-\rho_2 & = \frac{1}{2}+\eta+\chi+\frac{1}{2}\sqrt{1+4\mu^2 +\lambda +(\omega-m_{\phi})^2} \nonumber \\
& = \frac{1}{2}\left(1+|m_{\phi}|+\omega + \sqrt{1+m_{\psi}^2 +\lambda +(\omega-m_{\phi})^2}\right)  \\
\rho_2-\rho_1 &=  -\sqrt{1+m_{\psi}^2 +\lambda +(\omega-m_{\phi})^2}.
\end{eqnarray}
Checking these terms one by one we observe that only the three last terms can be equal to a negative integer.
In particular, focusing on the third and forth term and assuming that $\Re \omega>0$ for simplicity,
we get that the eigenvalues are
\begin{equation}
\label{lambda0}
\lambda=4n(1+n)-(2+4n)\left(\omega -|m_{\phi }|\right)-m_{\psi }{}^2,
\end{equation}
subject to the restriction
\begin{equation}
\label{energy_limit}
\Re \omega>1+|m_{\phi}|+2n.
\end{equation}
Note that since the condition for superradiance is $\Re \omega < m_{\phi} \Omega$,
the restriction (\ref{energy_limit}) means that indeed, superradiance is possible only for $\Omega>1$ and only for $m_{\phi}$s satisfying
\begin{equation}
\label{m_phi_restricion}
m_{\phi}> \frac{2n+1}{\Omega-1}.
\end{equation}
There is a second more technical restriction, in addition to (\ref{energy_limit}).
This comes from the fact that (\ref{zeroth}) also gives a higher order (non analytic) correction to the zeroth order eigenvalue we found. In particular, taking  $\gamma+\rho_1 = -n+\delta n$ and plugging this into the equation, we get the following: $\delta n \propto s^{\rho_1 - \rho_2}$ which implies that the correction to the eigenvalue itself will also be of this order.
In order to insure the validity of a perturbative expansion in $s$, up to first order, where this correction may be safely ignored, we require that $\Re(\rho_2 - \rho_1) \ge 2$.
This means that eigenvalues with
\begin{equation}
\label{validity_limit}
\Re \omega >3+|m_{\phi}|+2n
\end{equation}
are completely valid.
Taking into account the non analytical correction will make the entire computation harder since the expression $\rho_2 - \rho_1$ depends on $\omega$.
In this paper we perform the full computation without the correction, and discard those eigenvalues that do not respect this bound.

\subsubsection*{Subleading Correction}
\label{sub_leading_sec}
We would like to go beyond the zeroth order in $s=1-\frac{1}{a^2}$ in calculating the angular eigenvalue since in the zeroth order $\Omega_{\phi}=1$,
and we thus expect superradiance to be apparent only in the next-to-leading order.
To calculate the first order correction to the eigenvalue one needs to take into consideration the next order of the differential equations appearing in the two regions. The correction can be found separately in each region using perturbation theory and a comparison between them is needed to
ensure this is a valid solution in the entire segment. We describe this calculation in both regions in \ref{first_order_appendix}, and we only cite here the result (assuming $m_{\phi}\ge 0$ ):
\begin{eqnarray}
\label{lambda1}
\lambda _1&=-2+2 n+2 n^2+(1+2 n)(m_{\phi }- \omega)+ \nonumber \\
&\frac{\left(\left(\omega -m_{\phi }\right) m_{\phi }+(1+2n)\left(\omega -m_{\phi }\right)-2 n (1+n)\right)\left(m_{\psi }{}^2-4\right)}{\left(2 n-\omega +m_{\phi }\right) \left(2+2 n-\omega +m_{\phi }\right)}.
\end{eqnarray}
Notice that the last two terms, which are the only ones non-linear in $\omega$, cancel when $m_{\psi} = 2$.

We have implicitly assumed in taking the limit $s \rightarrow 0$ that all other parameters, and specifically $m_{\phi}$, are $O(s^0)$.
Since we have $\Omega_{\phi} = 1 + O(s)$ when $\alpha > 2$,
superradiance can occur only for $m_{\phi} \sim O(1/s)$,
as superradiant modes must also satisfy (\ref{m_phi_restricion}).
Thus, superradiance, and the instability associated with it, appear
to lie outside of the domain of validity of this approximation.
We discuss this issue in greater detail in section \ref{superradiant_result_sec}.

\subsection{The Radial Equation}
\label{radial_eq_sec}
The radial equation, depending only on the coordinate $x$, is given by
\begin{eqnarray}
\fl \frac{1}{x} \partial_x\left(xX\partial _xR(x)\right) +
\left( \frac{\left(\left(x^2+2m s_1^2\right)\left(-\omega +a m_{\phi }\right)+a m_{\phi } -a^2\omega \right)^2}{X}-\frac{a^2 m_{\psi }^2}{x^2}+\lambda \right)  R(x)=0.
\end{eqnarray}
Recall, from section \ref{extremal_limit_sec}, that upon making the coordinate transformation $z = x^2$,
the inner event horizon is located at $z = 0$ while the outer event horizon is at $z_2$.
These, along with the boundary of AdS at $z \rightarrow \infty$, are three of the
equation's regular singular points, the last one being located at
\begin{equation}
z_1=  -\frac{4 a +\left(a^2+1+z_2\right) \alpha }{\alpha }.
\end{equation}
Using the transformation $ r=\frac{z_2-z_1}{x^2-z_1}$
we move these to the canonical set of points $\{0,1,s,\infty\}$, with
\begin{equation}
s = 1 - \frac{z_2}{z_1}.
\end{equation}
One can now use the ansatz
\begin{equation}
\label{anzats_radial}
R(r)=r^{\nu }(r-1)^{\mu }\left(r-1+\frac{z_2}{z_1}\right)^{\rho }F(r)
\end{equation}
with
\begin{eqnarray}
\nu =  & 2 \\
\mu = & \frac{i \left(- \omega  \left(2 a +\alpha  \left(a^2+z_2\right)\right)+a m_{\phi } \left(2 a +\alpha  \left(1+z_2\right)\right)\right)}{2 \alpha  \sqrt{z_2} \left(-z_1+z_2\right)} \\
\rho= & \frac{a l m_{\psi }}{2 \sqrt{z_1} \sqrt{z_2}}
\end{eqnarray}
to transform the equation to the canonical form of Heun's equation (\ref{heun eq}).
The Heun parameters are then given by
\begin{eqnarray}
\label{Heun_prmtrs_radial}
\alpha_H= & \mu +\nu +\rho -i\sqrt{-\frac{z_2}{z_1}} \mu \\
\beta_H= &\mu +\nu +\rho +i\sqrt{-\frac{z_2}{z_1}} \mu\\
\gamma = & -1+2\nu =3 \\
\delta = & 1+2\mu \\
\epsilon = & 1+2\rho \\
q = &\frac{1}{4 \alpha  z_1 \sqrt{z_2}}\left(6 i a  \left(2 +a \alpha \right)) \omega +6 i a
\alpha  m_{\psi } \sqrt{-z_1}+a^2 \alpha  m_{\phi }^2 \sqrt{z_2}+ \right. \nonumber \\
& \left. \alpha  \left( \lambda + \omega ^2+16 z_1+6 i  \omega  \sqrt{z_2}-8 z_2\right) \sqrt{z_2}- \right. \nonumber \\
& \left. 2 i a m_{\phi } \left(3 l (2 a+\alpha )-i  \alpha  \omega  \sqrt{z_2}+3 \alpha  z_2\right)\right).
\end{eqnarray}

In order to solve this equation numerically, we use the Continued Fraction Method \cite{Gautschi1967},
which has first been applied to black hole physics in \cite{Leaver85,Leaver86}.
We first plug into the equation a series
solution, $F(r) = \sum_{n=0}^{\infty} a_n r^n$, satisfying the regularity
condition at the boundary of AdS, located at $r=0$.
This yields a three-term recurrence relation for $a_n$
\begin{equation}
f_k a_{k+1} + g_k a_k + h_k a_{k+1} = 0.
\end{equation}
The coefficients are
\begin{eqnarray}
\label{cont_frac}
f_k = &s (k+1)(k+\gamma)  \\
g_k = &-k \left[s \left(k+\gamma+\delta-1\right)+k+\gamma+\epsilon-1\right]-q  \\
h_k = & (k+\alpha_H-1)(k+\beta_H-1).
\end{eqnarray}
We also require regular boundary condition at the other end of the physical interval, $r=1$, which corresponds to the event horizon. This requirement is only satisfied by the minimal solution which in turn requires convergence of the following continued fraction \cite{Gautschi1967}
\begin{equation}
\label{quant_numer}
0=g_0-\cfrac{f_0\cdot h_1}{g_1-\cfrac{f_1\cdot h_2}{g_2-\cfrac{f_2\cdot h_3}{g_3-\cdot\cdot\cdot}}} \equiv g_0-\frac{f_0\cdot h_1}{g_1-} \cdot\frac{f_1\cdot h_2}{g_2-} \cdot\frac{f_2\cdot h_3}{g_3-}\cdot\cdot\cdot.
\end{equation}
This sets the quantization condition for $\omega$.

Truncating the continued fraction after some $p$ steps, we get a polynomial equation for $\omega$
which can be solved numerically.
Increasing the cutoff causes more quasi normal modes to be found, as well as better accuracy for the ones already found. This
is true as long as the precision of the computations is increased as well, so that numerical errors are minimized.

\section{Scalar Quasinormal Modes}
\label{QNM_res_section}
Using {\it Mathematica} we carried out the numerical procedure for various choices of the parameters. In order to achieve numerical stability, a large cutoff of the continued fraction was used, often over $140$ terms, in high precision floating point arithmetics. Below we present an example of the typical results obtained using the numerical procedure.

\begin{figure}[!ht]
    \includegraphics[width=0.65 \linewidth,scale=0.5]{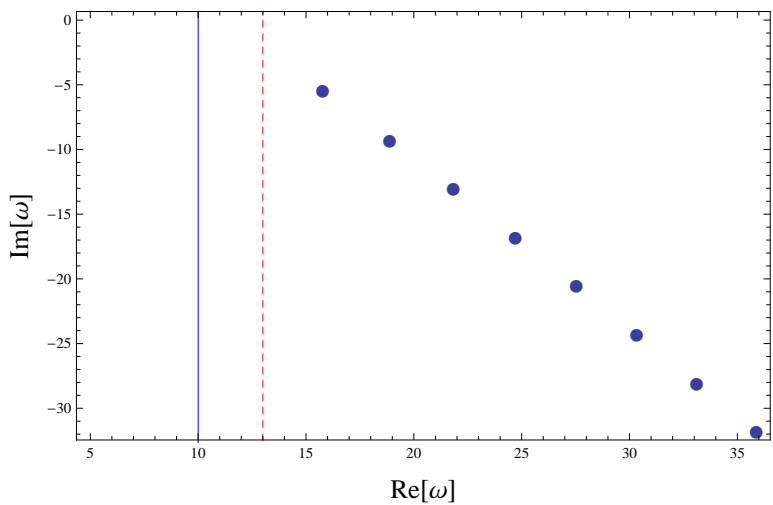}
    \centering
    \caption{First 8 QNMs for $\alpha=2$, $a=0.99$, $z_2=0.2$, $n=1$, $m_{\phi}=7$, $m_{\psi}=2$. The continuous blue line denotes the energy bound (\ref{energy_limit}), the dashed red line denotes the bound for perturbation validity (\ref{validity_limit})  }
    \label{stable_modes_figure}
\end{figure}

As expected, for small $m_{\phi}$s, we did not find any instabilities using a
variety of values for $\alpha$, $m_{\phi}$, $m_{\psi}$, $z_2$ and $n$.
As an example, Figure \ref{stable_modes_figure} displays the first quasinormal modes for the following parameters
$\alpha = 2$, $a = 0.99$, $z_2 = 0.2$, $m_{\phi} = 7$, $m_{\psi} = 2$, $n=1$.
\FloatBarrier
The relative numerical error resulting from the cutoff of the continued fraction is less than 0.0001.

\subsection{Observing Superradiant Instabilities}
\label{superradiant_result_sec}
In order to observe superradiant modes, we require $\omega < m_{\phi} \Omega_{\phi}$, but we also know that the approximate solution is valid only for $\omega > m_{\phi} + 1$. Thus, we must have $m_{\phi} > \frac{1}{\Omega_{\phi} - 1} \propto \frac{1}{|s|}$. Having such a large parameter in the problem
turns the angular equation into the more complicated Confluent Heun Equation.
However, naive computations using the numerical method indeed show the known
superradiant instabilities, as found in other types of black holes \cite{Cardoso2004,Aliev2008, Cardoso2004a,Cardoso2006,Cardoso2005, Konoplya2011}
As an example of this superradiant instability,
we display the first quasinormal modes for the following parameters
$\alpha=2.1$, $a=0.9999$, $z_2=0.01$ ,$n=0$, $m_{\phi}=10^7$, $m_{\psi}=2$.
In this case, we have $\Omega_{\phi} - 1 \approx 1.9 \times 10^{-6}$.
\begin{figure}[!ht]
    \includegraphics[width=0.65 \linewidth,scale=0.5]{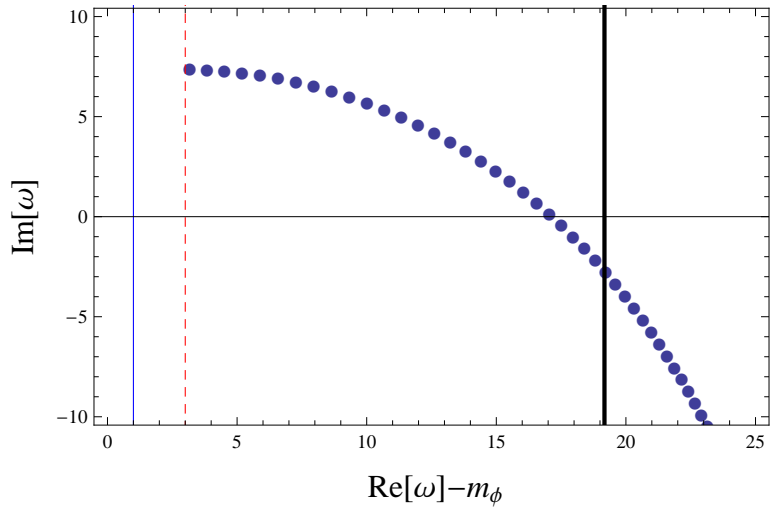}
    \centering
    \caption{QNMs for $\alpha=2.1$, $a=0.9999$, $z_2=0.01$ ,$n=0$, $m_{\phi}=10^7$, $m_{\psi}=2$. The continuous blue line denotes the energy bound (\ref{energy_limit})
    and the dashed red line denotes the bound for perturbation validity (\ref{validity_limit}).
    The thick black line shows the expected superradiance threshold.}
    \label{superradiant_figure}
\end{figure}
We can estimate the errors to this numerical computation by increasing the cutoff on the continued fraction, and checking the change in the eigenfrequencies. In this case, varying the cutoff from $p=100$ to $p=140$ shows that the real part of the frequencies is correct to about $10^{-9}$, while their imaginary part is less accurate. Far enough from the crossing point between positive and negative imaginary part we have an error of approximately $0.05$, while close enough to the superradiance threshold (depicted by the black line in Figure \ref{superradiant_figure}), errors grow up to about $1$. This explains why the actual crossing point is not precisely at the superradiance threshold.

The well-known phenomenon of superradiance thus occurs where expected. The reason that the scaling $m_{\phi} \sim \frac{1}{|s|}$ does not mask it, is that even with this scaling, we still expect $\left(\omega -m_{\phi }\right)$ to be finite without introducing new singularities.
Indeed, if we search for eigenmodes having
$\omega = m_{\phi} + \Delta$ with $\Delta \sim O(1)$,
$\lambda_1$ is still finite, with the following exception.
There is one term that, for $m_{\psi} \neq 2$, is of order $O(1/|s|)$
(the term is absent for $m_{\psi} = 2$, which is depicted in Figure \ref{superradiant_figure}).
This term changes the zeroth order term in the expansion of $\lambda$, but it remains finite.
Thus, the result for $\lambda$, obtained
from the next-to-leading order of the asymptotic matching procedure,
is finite in the limit $s \rightarrow 0$.
If there are no further corrections to the zeroth order term coming from higher order
terms when using this scaling, the results obtained here would be correct at leading order.
We have not proven that this happens, but the compatability of our results with the expected superradiance threshold support it.

\section{Thermodynamic Stability}
\label{thermodynamic_section}
As previously mentioned, thermodynamic instabilities
manifest themselves as non-positive modes of the thermodynamic hessian
\begin{equation}
H_{m n} = -\frac{\partial^2 S}{\partial x^m \partial x^n};\ \ x^m = (E, J_{\phi} \equiv J, Q)
\end{equation}
In order to compute the Hessian, we first express the entropy using these variables, by inverting (\ref{charges}).
Then, after computing the Hessian, we would like to express the result
using the parameters $a$, $\alpha$ and $z_2$.
However, it is more insightful to
express the Hessian in terms of ${\alpha, \delta, Q}$ where $\alpha$ and $\delta$ are defined in the following way
\begin{eqnarray}
\alpha &= \frac{J}{Q^2} \\
\delta &= \frac{E-J}{Q},
\end{eqnarray}
and make a large-charge expansion, taking the limit $Q\rightarrow\infty$. This limit is equivalent to the limit $a \rightarrow 1$ we have considered thus far and
in taking it we must keep $\alpha,\delta \sim O(Q^0)$.
At leading order in $Q$ it is easily verified that the extremality is achieved for $\delta$ taking the value
$\bar{\delta} = \frac{2+\alpha}{\sqrt{2\alpha}}$,
We thus substitute $\delta = \bar{\delta} + \delta_0$, so that
states above extremality correspond to values of $\delta_0 > 0$.
The Hessian at leading order is given by
\begin{eqnarray}
\label{hessian}
H = \frac{\pi}{ 2^{5/4} Q (\sqrt{\alpha} \delta_0)^{3/2}}
\left(
\begin{array}{ccc}
\alpha & -\alpha & -2 \sqrt{2\alpha} \\
-\alpha & \alpha & 2 \sqrt{2\alpha} \\
-2 \sqrt{2\alpha} & 2 \sqrt{2\alpha} & 4\sqrt{2\alpha}+8 \delta_0
\end{array}
\right).
\end{eqnarray}
The eigenvalues of the hessian are
\begin{eqnarray}
\lambda_1 &= 0  \\
\lambda_2 &= \frac{4 + \alpha + 2\sqrt{2\alpha} + \sqrt{(4 + \alpha)^2 -4\sqrt{2\alpha}(\alpha - 4) \delta_0 + 8\alpha \delta_0^2}}
{2^{5/4} \alpha^{9/4} Q \delta_0^{3/2}}  \\
\lambda_2 &= \frac{4 + \alpha + 2\sqrt{2\alpha} - \sqrt{(4 + \alpha)^2 -4\sqrt{2\alpha}(\alpha - 4) \delta_0 + 8\alpha \delta_0^2}}
{2^{5/4} \alpha^{9/4} Q \delta_0^{3/2}}
\end{eqnarray}
Thus, at leading order the hessian always has a zero mode, and we must go to next-to-leading order in $Q$.
Since we know $\lambda_2$ and $\lambda_3$ to leading order,
the determinant
\begin{equation}
\left| H \right| = \frac{2^{5/4}-2^{1/4} \alpha +2^{3/4} \sqrt{\alpha } \delta_0}{8 \alpha ^{9/4} \delta _0^{5/2} Q^5} + O\left(\frac{1}{Q^6}\right).
\end{equation}
allows us to extract the leading order contribution to $\lambda_1$, and we find:
\begin{eqnarray}
\label{eigenvalues}
\lambda_1 &= \frac{2-\alpha +\sqrt{2} \sqrt{\alpha } \delta _0}{ 2^{15/4} Q^3 \alpha ^{9/4} \sqrt{\delta _0}} =\frac{1-\Omega _{\phi }}{2^{\frac{9}{4}}Q^2 \alpha ^{3/4} \sqrt{\delta _0}}  \\
\lambda_2 &= \frac{4 + \alpha + 2\sqrt{2\alpha} + \sqrt{(4 + \alpha)^2 -4\sqrt{2\alpha}(\alpha - 4) \delta_0 + 8\alpha \delta_0^2}}
{2^{5/4} \alpha^{9/4} Q \delta_0^{3/2}} > 0 \\
\lambda_2 &= \frac{4 + \alpha + 2\sqrt{2\alpha} - \sqrt{(4 + \alpha)^2 -4\sqrt{2\alpha}(\alpha - 4) \delta_0 + 8\alpha \delta_0^2}}
{2^{5/4} \alpha^{9/4} Q \delta_0^{3/2}} > 0
\end{eqnarray}
where when expressing $\lambda_1$ in terms of $\Omega _{\phi}$ we used the expansion of $\Omega _{\phi }$ up to order $O(\frac{1}{Q^2})$.

The result (\ref{eigenvalues}) reflects the fact that the black hole is thermodynamically unstable when $\Omega _{\phi }>1$. This is the same regime where superradiant instability has been shown to occur in the previous section, a fact which might be related to a Gubser-Mitra-like conjecture \cite{Gubser2001} for rotating black holes, though there are some differences, as discussed in the next section.
The eigenvector corresponding to this instability is given by $(1,1,0)$,
when written in terms of $(\delta E, \delta J, \delta Q)$.

In order to describe the onset of the instability in terms of the charges $E,J,Q$,
let us use them to re-express $\Omega_{\phi }$:
\begin{equation}
\Omega _{\phi }=1+\frac{-E+\sqrt{2} \sqrt{J}+J}{2 J}+\left(\frac{1}{Q^2}\right).
\end{equation}
Thus, the stability would appear for $E-J < 2\sqrt{J}$. In particular, for very large energies there is no instability while for very large angular momentum instability is guaranteed.

It is also interesting to examine the temperature in terms of the charges
\begin{equation}
T=\frac{\sqrt{-J \left(-2 E \sqrt{J}+\sqrt{2} J+2 J^{3/2}+2 \sqrt{2} Q^2\right)}}{2^{1/4} J \pi }+O\left(\frac{1}{Q}\right).
\end{equation}
Therefore, as we could have also concluded from (\ref{hessian}),
the temperature is an increasing function of energy at fixed $J$ and a decreasing function of the angular momentum at fixed $E$.

For completeness, we express $\delta$ in terms of the parameters $a,\alpha,z_2$
\begin{eqnarray}
\delta & = \frac{4a^2(3+a)+4a\left(2+a+a^2+(3+a)z_2\right) \alpha +(3+a)\left(1+z_2\right)\left(a^2+z_2\right) \alpha ^2}{2\sqrt{2}(1+a)\sqrt{a \alpha \left(2a+\alpha +z_2 \alpha \right)\left(2a+\left(a^2+z_2\right) \alpha \right)}} \nonumber \\
& = \bar{\delta}+\frac{\sqrt{\alpha } z_2}{\sqrt{2}}+O(1-a)
\end{eqnarray}

\section{Summary and Discussion}
\label{discussion}
We have analyzed the stability of a three parameter family of black holes
charged under two of the $U(1)$s in the cartan of $SO(6)$, within the consistent truncation of  type $IIB$ SUGRA on $AdS_5 \times S^5$. The specific black holes that we studied had two equal charges (denoted by $Q$).
The black holes also have non-zero angular momentum around
one of the two independent rotation
planes of $AdS_5$, denoted by $J_{\phi}$.
We have defined $\alpha \equiv N^2 J_{\phi} / Q^2$ - a quantity measuring
the ratio between the angular momentum  and the
charge squared. For this class of black holes, as we approach the extremal limit, the horizon shrinks to zero size.
Such black holes have been dubbed Extremal Vanishing Horizon (EVH), and their properties have been studied in \cite{Sheikh-Jabbari2011,Boer2011,Johnstone2013}.
For a generic point on the horizon, the extremal near horizon limit is pinched $AdS_3$. In contrast, on the plane of rotation the horizon merges with a singularity as we take the extremal limit, and no universal near horizon exists at the level of the geometry.
The value $\alpha=2$ is distinguished in the sense that it satisfies, at extremality, a SUSY BPS bound.

We carried out a stability analysis in the limit of $J_{\phi} \rightarrow \infty$
for fixed $\alpha$. In the notation of the solution given in (\ref{metric2}), this limit corresponds to the limit in which the rotation parameter  $a \rightarrow 1$. We have checked for both local thermodynamic instabilities and classical instability,
analyzed using linear perturbations analysis. The classical stability analysis was carried out for a scalar field, after identifying an appropriate scalar whose addition still forms a consistent truncation of type $IIB$ SUGRA on $AdS_5 \times S^5$.

We found evidence for both types of instabilities when $\Omega_{\phi} > 1$,
where $\Omega_{\phi}$ is the angular velocity of the black hole with respect
to a non-rotating frame at the boundary of AdS.
This is also the threshold predicted in \cite{Hawking2000} for superradiant instability,
induced by the asymptotically AdS nature of spacetime. Indeed, the unstable modes are precisely
those which are superradiant.
In the near extremal limit this condition translates to $\alpha > 2$.

It is important to note that, in the limit $a \rightarrow 1$,
the angular velocity takes the form $\Omega_{\phi} = 1 + O(1-a)$
and one must therefore examine the sub-leading term to see if $\Omega_{\phi} > 1$ or not.
In addition this form of $\Omega_{\phi}$ implies that a superradiant instability can appear only for modes of the scalar having
very large angular momentum, $m_{\phi} \sim 1/(1-a)$.

It would be interesting to find out what would be the endpoint
of these instabilities.  Unfortunately, knowing there is a
thermodynamic instability does not allow us to extract much information
regarding the final state, and the classical linear stability analysis is obviously insufficient for such purposes,
and should be replaced by a full non-linear analysis.

A natural question to ask at this point is whether the two instabilities are related. It has been shown numerically that
for $AdS_4$ Reissner Nordstrom (RN) black holes, in the brane limit,
local thermodynamic stability coincides with dynamical stability \cite{Gubser2000,Gubser2001}.
This phenomenon was shown to persist also for
black holes with horizon radius comparable to $AdS$ radius.
The authors further suggested that it is a
generic phenomenon, in what came to be known as
the Gubser-Mitra conjecture \cite{Gubser2001}.
A part of this conjecture, which states that black branes are unstable if, and only if,
they are constructed from a thermodynamically unstable black hole (at lower dimension),
has recently been proven in \cite{Hollands2012}.
This work also established a close connection between
dynamical and thermodynamical stability of black holes,
but unfortunately it has not yet
been extended to Anti-de Sitter space. It also does not apply to our case, since the superradiance instabilities that we find are at short wave length in the $J\rightarrow\infty$ limit.

In the context of fast-rotating black holes, which we are discussing here,
intuition can be obtained from arguments
originating in the study of higher dimensional asymptotically flat black holes.
It has been argued \cite{Emparan2003} that black holes in
$D \ge 6$ with one non-vanishing rotation parameter are prone to ultra-spinning instabilities.
Black holes are ultra-spinning when there is no upper bound on the angular momentum for
fixed mass. For large enough angular momentum, the event horizon becomes quasi-extended,
resembling a black brane.
The observed instabilities, which are purely gravitational, are in fact Gregory-Laflamme (GL)
type instabilities \cite{Gregory1994}, causing fragmentation (or clumping)
of the extended black hole. As mentioned above, for black branes, such instabilities can
be identified by studying the local thermodynamic instabilities. This can be
understood intuitively by arguing that the clumped black holes are
entropically favored when a local thermodynamic instability exists.

As mentioned in \cite{Emparan2003}, the above argument cannot be made precise in $D=5$ since
the black holes in that case have no ultra-spinning regime,
and the black brane limit is not defined since there are no black two-branes.
However, fast-rotating black holes in $D=5$ asymptotically flat spacetime
can still become very much 'pancaked', as is the case for the black holes studied in this paper.
It has been conjectured \cite{Emparan2002} that
such black holes become unstable before extremality,
with the endpoint of the instability being a black ring.
This black ring is thermodynamically favored, so that a
thermodynamic instability is expected for the black hole. In
\cite{Shibata2009} a classical gravitational
non-axisymmetric instability was found which approximately matches
the instability region predicted by thermodynamic
arguments. Whether this instability also occurs in $AdS_5$ black
holes is, to the best of our knowledge, unknown.
We have not yet performed a gravitational stability analysis,
so it is not clear if such an instability exists, but it is obviously a viable option.

However, this is not the only possible outcome for the black hole evolution.
Superradiant instability, which is mostly absent from the discussions in the above cases,
occurs also at zero temperature (i.e. in extremal black holes).
It can be interpreted within string microscopic models of
black holes as a quantum effect \cite{Dias2007a}.
At zero temperature, superradiance is taken to be spontaneous emission,
while at non-zero temperature, the thermal radiation is thrown into the mix.
It is possible that the instability observed in the thermodynamic analysis reflects this as well.
The stable configuration, if this is the dominant mechanism of decay, is most likely of the type predicted in \cite{Cardoso2006} for four dimensional AdS black holes.
It is a slightly modified $\Omega_{\phi} = 1$ black hole co-existing with some
outside scalar hair and radiation. A scalar field instability of small charged and rotating black holes in $AdS_5$ was shown numerically
to result in a black hole with scalar hair \cite{Brihaye2011} (also \cite{Basu2010}). Hairy black holes were also shown to occur
in $AdS_5 \times S^5$ \cite{Bhattacharyya2010}.
In our case, we expect that the final result would be an $\alpha=2$ core surrounded by hair.
Of course the analysis for the black holes under consideration here is further complicated by the fact that they have a
non-trivial configuration of both the gauge fields and scalars.

The black holes which are the focus of this paper have been
conjectured in \cite{Berkooz2012} to correspond to Fermi surface-like operators
in the $PSU(1,1|2)$ sector of the dual conformal field theory, ${\cal N}=4$ $SU(N)$ $SYM$.
Understanding the final state evolving from the unstable black holes
can help one study the dual CFT vacuum configuration
for $\alpha > 2$. Conversely, if this instability can be observed on the field theory side,
perhaps it would hint at the end-point of the gravity theory's instability.

The classical instabilities observed in this paper are for neutral fields,
carrying no charges out of the black holes. It would be interesting
to see if the black hole can also emit some of its charge to the outside. This can be studied
by either adding charged scalars to the consistent truncation, or using charged vector fields, using the truncation to $SU(2) \times U(1)$ instead of $U(1)^3$ \cite{Lu2000}.
Another possibility would be to compute Fermionic two-point functions, which,
in addition to charge emission, can also be used to observe Fermi surfaces in the dual CFT.
Such fermi surfaces have been observed for the black-brane analogs of these black holes \cite{DeWolfe2012a}.

\ack
We are grateful to J.~Simon for many suggestions and enlightening conversations,
and collaboration at early stages of this project.
This work was supported in part by the Israel-U.S. Binational Science Foundation, by the Israel Science Foundation
(grant number 1665/10),  by the German-Israeli Foundation (GIF) for Scientific
Research and Development (grant number I-1-038-47.7/2009) and by a Minerva grant. 
\newpage

\appendix
\section{Elimination of additional possible quantization conditions for \texorpdfstring{$\lambda$}{lambda}}
\label{El_angular}
We need to consider whether the relation $\rho_2-\rho_1=-n+O(s)$ might lead to some quantization condition. This was treated rather generally in the Appendix of our previous work, up to a point which depended on the parameters of the equation. We will treat this point here, using the notations of that Appendix.
The non generic case to consider is when $a=-m$ or $b=-m$  and $n<m$ for both the far and the near region in which case the expansions in $s$ leading to (\ref{zeroth}) are valid. $a$ and $b$ for the far region are defined by
\begin{eqnarray}
\label{last_case}
a_{far}= & \beta_H+\rho_2 \nonumber \\
b_{far}=& \alpha_H+\rho_2
\end{eqnarray}
while for the near region they are
\begin{eqnarray}
a_{near}= & -\rho_1 \nonumber \\
b_{near}=& -\rho_1-\gamma+1 =  -|m_{\phi}|-\rho_1
\end{eqnarray}
Now we can rewrite (\ref{zeroth}) as
\begin{equation}
\frac{\Gamma^2 \left[n\right] \Gamma \left[\gamma +\rho _2\right]\Gamma \left[a_{near}\right]\Gamma \left[a_{far}\right]\Gamma \left[b_{far}\right]}{\Gamma \left[\alpha_H +\rho _1\right]\Gamma \left[\beta_H +\rho _1\right] \Gamma^2 \left[-n\right]\Gamma \left[\gamma +\rho _1\right]\Gamma \left[-\rho _2\right]}=\left(-s\right)^{2n}
\end{equation}
We observe that for this equality to be fulfilled we need to have at least one more divergent gamma function in the denominator than in the numerator of the LHS since the RHS goes to zero. We will show this doesn't happen, assuming throughout our considerations that the only divergent term in the denominator is $\left(\Gamma \left[-n\right]\right){}^2$ since the other cases were already treated.
Thus, we need to check only the cases where there is only one divergent gamma function in the numerator \\
The only such case is that one of  $a_{far}$, $b_{far}$ is equal to $-m_1,-m_2>-n$ and $b_{near}$ also fulfills this requirement but $a_{near}$ doesn't fulfill it. This means (1) $-m_{\phi}-\rho_1<0$ and (2) $|m_{\phi}+\rho_1|=m_1<n$. From (2) $\rho_1$ is an integer. If $-\rho_1<0$ then from (2) $|\rho_1|<n$ so $a_{near}$ fulfills the conditions contrary to our assumptions. If $\rho_1=0$ then $\Gamma[-\rho_1]$ is divergent so the LHS is finite and incompatible to the RHS. Therefore $\rho_1<0$ and  from (1) $-m_{\phi}+|\rho_1|<0$ $\Rightarrow$ $|\rho_1|< m_{\phi}$. Now we can rewrite (2) as $m_{\phi}-|\rho_1|<n$ so $m_{\phi}<n$. Recalling that $\rho_1=n+\rho_2$ we get that $\rho_2$ is an integer and using the fact that $\rho_1<0$ we get that $\rho_2<-n$. Now we shall prove that the in this case $\Gamma[\gamma+\rho_2]$ is divergent.
\begin{equation}
\gamma+\rho_2=1+m_{\phi}+\rho_2 < 1+m_{\phi}-n < 1
\end{equation}
since both $\gamma$ and $\rho_2$ are integers we got that $\gamma+\rho_2$ is an integer at most equal to zero and this ends our proof.
\normalsize
\section{First order correction to the \texorpdfstring{$\lambda$}{lambda} in the \texorpdfstring{$a \rightarrow l$}{a goes to l} case}
\label{first_order_appendix}
\subsection*{The far region}
In the zeroth order we had the equation
\begin{equation}
\label{zeroth_eq_far}
D_0 {\psi}_n(u) =  q_0(n){\psi}_n(u) 
\end{equation}
with the differential operator 
\begin{equation}
D_0   =  u\left[(u-1) u \partial_u^2+((u-1) (1+\alpha_H +\beta_H )+\delta ) \partial_u+ \alpha_H  \beta_H \right]
\end{equation}
and the solution was given by
\begin{eqnarray}
{\psi}_n^0 &=  u^{\rho_1} F_{1}[\alpha_H+\rho_1,\beta_H+\rho_1,\delta,1-u] \nonumber \\
& = u^{-\gamma-n} F_{1}[\alpha_H-\gamma-n,\beta_H-\gamma-n,\delta,1-u]
\end{eqnarray}
after using the quantization condition from the matching.\\
Thus, taking $F(u)={\psi}_n^0+{\psi}_n^1$ as the solution one gets the equation up to first order in $s$
\begin{equation}
(D_0+s D_1)({\psi}_n^0+s {\psi}_n^1)=(q_0(n)+s q_1)({\psi}_n^0+s {\psi}_n^1),
\end{equation}
which gives the following equation for the unknowns ${\psi}_n^1$ and $q_1$
\begin{equation}
D_0 {\psi}_n^1-q_0(n){\psi}_n^1=q_1 {\psi}_n^0-D_1{\psi}_n^0
\end{equation}
One can actually rewrite the differential operator of the full equation as $D_0+c D_1$ where
\begin{eqnarray}
D_1 & = u(1-u)\partial_u^2+((1+\alpha_H+\beta_H)(1-u)-\delta)\partial_u-\epsilon (1-r)\partial_u \\
& = -\frac{1}{u}D_0-\epsilon (1-u)D_r+\alpha_H \beta_H
\end{eqnarray}
Then, using the following recurrence relations
\begin{equation}
\frac{1}{u} {\psi}_n^0(u)=\sum_{i=-1}^{i=1}k_{n+i}{\psi}_{n+i}^0
\end{equation}
with
\begin{eqnarray}
k_{n+1} &= & \frac{\left(\alpha_H +\rho _2\right)\left(\beta_H +\rho _2\right)}{\left(\rho _1-\rho _2\right)\left(\rho _1-\rho _2-1\right)}\\
k_n & = & \frac{-1+\beta_H +\rho _1+\rho _2-2 \rho _1 \rho _2-\beta_H \left(\rho _1+\rho _2\right)}{\left(\rho _1-\rho _2-1\right)\left(\rho _1-\rho _2+1\right)} - \nonumber \\ 
&& \frac{\alpha_H  \left(-1+2 \beta_H +\rho _1+\rho _2\right)}{\left(\rho _1-\rho _2-1\right)\left(\rho _1-\rho _2+1\right)}\\
k_{n-1} &= & \frac{\left(\alpha_H +\rho _1\right)\left(\beta_H +\rho _1\right)}{\left(\rho _1-\rho _2\right)\left(\rho _1-\rho _2+1\right)}
\end{eqnarray}
and
\begin{equation}
(u-1)\partial_u {\psi}_n(u)=\sum_{i=-1}^{i=1}p_{n+i}{\psi}_{n+i}^0
\end{equation}
with
\begin{eqnarray}
p_{n-1} & =  -\frac{\left(\alpha_H +\rho _1\right) \left(\beta_H +\rho _1\right) \rho _2}{\left(\rho _1-\rho _2\right) \left(1+\rho _1-\rho _2\right)} \\
p_n & =  \frac{\rho _1\rho _2\left(\rho _1+\rho _2+2\beta_H -1\right)+\alpha_H \left(2\rho _1\rho _2+\beta_H \left(1+\rho _1+\rho _2\right)\right)}{\left(\rho _1-\rho _2-1\right)\left(\rho _1-\rho _2+1\right)}\\
p_{n+1} & =  -\frac{\left(\alpha_H +\rho _2\right) \left(\beta_H +\rho _2\right) \rho _1}{\left(\rho _1-\rho _2\right) \left(-1+\rho _1-\rho _2\right)}
\end{eqnarray}
So finally we can write
\begin{equation}
(D_0-q_0(n)){\psi}_n^1=q_1 {\psi}_n^0-\sum_{i=-1}^{i=+1} (\epsilon p_{n+i}-k_{n+i}q_0(n)) {\psi}_{n+i}^0
\end{equation}
We must use a solution of the form ${\psi}_n^1=\sum_{i=-1}^{i=+1}A_{n+i}u_{n+i}^0$. Since ${\psi}_n^0$ are eigenfunctions of $D_0$ with different eigenvalues they are linearly independent and we get an algebraic equation for the coefficients $A_j$. Examining  the coefficients of $u_{n}$ we get an expression for $q_1$
\begin{equation}
q_1=\alpha_H \beta_H +\epsilon p_n - q_0(n)k_n
\end{equation}
This means we got the next order of the eigenvalue without using the matching of the two regions. To make sure this makes sense we should perform the same procedure in the near region and compare the two corrections.
\subsection*{The near region}
The zeroth order equation is
\begin{equation}
\label{zeroth_eq_near}
\tilde{D_0} {\chi}_n(\xi)   =  q_0(n){\chi}_n(\xi) 
\end{equation}
with the differential operator 
\begin{equation}
\tilde{D_0} = -\xi  (1+\xi )\partial_{\xi}{}^2-(\gamma +(1+\alpha_H +\beta_H -\delta ) \xi )\partial_{\xi}
\end{equation}
and the solution was given by 
\begin{eqnarray}
{\chi}_n & = F_1\left(-\rho_1,-\rho_2,\gamma ;-\xi \right) \nonumber \\
& = F_1\left(\gamma+n,\epsilon-1-n,\gamma ;-\xi \right).
\end{eqnarray}
The next order equation is
\begin{equation}
\label{1order near}
\tilde{D_0}{\chi}_n^1-q_0(n){\chi}_n^1=q_1 {\chi}_n^0-\tilde{D_1} {\chi}_n^0
\end{equation}
with
\begin{eqnarray}
\tilde{D_1} & =  -\xi ^2 (1+\xi ) D_{\xi }{}^2- \xi  (\gamma +\delta +(1+\alpha_H +\beta_H ) \xi ) D_{\xi }- \alpha_H  \beta_H  \xi  \nonumber \\
& = \xi \tilde{D_0}-\delta \xi (\xi+1)\partial_{\xi}- \alpha_H \beta_H \xi
\end{eqnarray}
One can express
\begin{equation}
\xi {\chi}_n=\sum_{i=-1}^{i=+1}f_{n+i} {\chi}_{n+i}
\end{equation}
with
\begin{eqnarray}
& f_{n-1}=\frac{-\left(\rho _1+\gamma \right)\rho _2}{\left(\rho _2-\rho _1-1\right)\left(\rho _2-\rho _1\right)}\\
& f_n=\frac{2\rho _1 \rho _2+\gamma \left(1+\rho _1+\rho _2\right)}{\left(\rho _2-\rho _1+1\right)\left(\rho _2-\rho _1-1\right)}\\
&f_{n+1}=\frac{-\left(\rho _2+\gamma \right)\rho _1}{\left(\rho _2-\rho _1+1\right)\left(\rho _2-\rho _1\right)}
\end{eqnarray}
and
\begin{equation}
\xi (\xi +1)D_{\xi } {\chi}_n=\sum_{i=-1}^{i=+1}e_{n+i} {\chi}_{n+i}
\end{equation}
with
\begin{eqnarray}
& e_{n-1}=\frac{-\rho _1\rho _2\left(\gamma +\rho _1\right)}{\left(\rho _2-\rho _1\right)\left(\rho _2-\rho _1-1\right)}\\
& e_n=\frac{ \rho _1\rho _2\left(\rho _1+\rho _2+2\gamma -1\right)}{\left(\rho _2-\rho _1-1\right)\left(\rho _2-\rho _1+1\right)} \\
& e_{n+1}=\frac{-\rho _1\rho _2\left(\gamma +\rho _2\right)}{\left(\rho _2-\rho _1\right)\left(\rho _2-\rho _1+1\right)}
\end{eqnarray}
Using these expressions and writing ${\chi}_n^1$ as a linear combination of ${\chi}_{n+1}^0, with -1<i<1$, in (\ref{1order near}), 
one can proceed in the same manner as in the far region and obtain
\begin{equation}
\label{eq_q1}
q_1=(q_0(n)-\alpha_H \beta_H)f_n -\delta e_n.
\end{equation}
Indeed, comparing the two expressions for $q_1$ from the far and near region, using $q_0=\rho_1 \rho_2$, one gets an equality.
We can now proceed to compute the first order correction for $\lambda$ from the expression for $q_1$. We have $q=q_0+ s q_1$ with
\begin{equation}
q_0=\frac{1}{4} \left(2 m_{\phi }-m_{\psi }^2-\lambda _0+2\omega +4m_{\phi } \omega \right)
\end{equation}
and
\begin{equation}
q_1=\frac{1}{4} \left(2 m_{\phi }+m_{\phi }{}^2+2 m_{\psi }+2 m_{\phi }m_{\psi }+m_{\psi }{}^2+\lambda _0- m_{\phi } \omega \right)-\frac{1}{4}\lambda_1.
\end{equation}
Plugging this into (\ref{eq_q1}) and solving for $\lambda_1$ we are left with
\begin{eqnarray}
\lambda _1&=-2+2 n+2 n^2-(1+2 n) \omega +m_{\phi }+2 n m_{\phi }+ \nonumber \\
&\frac{\left(-4+m_{\psi }{}^2\right) n \left(n+m_{\phi }\right)}{2 n-\omega +m_{\phi }}-\frac{\left(-4+m_{\psi }{}^2\right) (1+n) \left(1+n+m_{\phi }\right)}{2+2 n-\omega +m_{\phi }}.
\end{eqnarray}

\section*{References}
\bibliographystyle{JHEP}
\bibliography{blackholes}	

\end{document}